\documentstyle[epsfig]{mn}

\newif\ifAMStwofonts

\ifoldfss
  \ifCUPmtlplainloaded \else
    \NewTextAlphabet{textbfit} {cmbxti10} {}
    \NewTextAlphabet{textbfss} {cmssbx10} {}
    \NewMathAlphabet{mathbfit} {cmbxti10} {} 
    \NewMathAlphabet{mathbfss} {cmssbx10} {} 
  \fi
  \ifAMStwofonts
    \ifCUPmtlplainloaded \else
      \NewSymbolFont{upmath} {eurm10}
      \NewSymbolFont{AMSa} {msam10}
      \NewMathSymbol{\upi}     {0}{upmath}{19}
      \NewMathSymbol{\umu}     {0}{upmath}{16}
      \NewMathSymbol{\upartial}{0}{upmath}{40}
      \NewMathSymbol{\leqslant}{3}{AMSa}{36}
      \NewMathSymbol{\geqslant}{3}{AMSa}{3E}

       \let\le=\leqslant
       
    \fi
  \fi
\fi 

\ifnfssone
  \newmathalphabet{\mathit}
  \addtoversion{normal}{\mathit}{cmr}{m}{it}
  \addtoversion{bold}{\mathit}{cmr}{bx}{it}
  \newmathalphabet{\mathbfit} 
  \addtoversion{normal}{\mathbfit}{cmr}{bx}{it}
  \addtoversion{bold}{\mathbfit}{cmr}{bx}{it}
  \newmathalphabet{\mathbfss} 
  \addtoversion{normal}{\mathbfss}{cmss}{bx}{n}
  \addtoversion{bold}{\mathbfss}{cmss}{bx}{n}
  \ifAMStwofonts
    \ifCUPmtlplainloaded \else
      \UseAMStwoboldmath
      \makeatletter
      \new@mathgroup\upmath@group
      \define@mathgroup\mv@normal\upmath@group{eur}{m}{n}
      \define@mathgroup\mv@bold\upmath@group{eur}{b}{n}
      \edef\UPM{\hexnumber\upmath@group}
      \new@mathgroup\amsa@group
      \define@mathgroup\mv@normal\amsa@group{msa}{m}{n}
      \define@mathgroup\mv@bold\amsa@group{msa}{m}{n}
      \edef\AMSa{\hexnumber\amsa@group}
      \makeatother
      \mathchardef\upi="0\UPM19
      \mathchardef\umu="0\UPM16
      \mathchardef\upartial="0\UPM40
      \mathchardef\leqslant="3\AMSa36
      \mathchardef\geqslant="3\AMSa3E

       \let\le=\leqslant

    \fi
  \fi
\fi 

\ifnfsstwo
  \DeclareMathAlphabet{\mathbfit}{OT1}{cmr}{bx}{it}
  \SetMathAlphabet\mathbfit{bold}{OT1}{cmr}{bx}{it}
  \DeclareMathAlphabet{\mathbfss}{OT1}{cmss}{bx}{n}
  \SetMathAlphabet\mathbfss{bold}{OT1}{cmss}{bx}{n}
  \ifAMStwofonts
    \ifCUPmtlplainloaded \else
      \DeclareSymbolFont{UPM}{U}{eur}{m}{n}
      \SetSymbolFont{UPM}{bold}{U}{eur}{b}{n}
      \DeclareSymbolFont{AMSa}{U}{msa}{m}{n}
      \DeclareMathSymbol{\upi}{0}{UPM}{"19}
      \DeclareMathSymbol{\umu}{0}{UPM}{"16}
      \DeclareMathSymbol{\upartial}{0}{UPM}{"40}
      \DeclareMathSymbol{\leqslant}{3}{AMSa}{"36}
      \DeclareMathSymbol{\geqslant}{3}{AMSa}{"3E}

       \let\le=\leqslant

    \fi
  \fi
\fi 

\ifCUPmtlplainloaded \else
  \ifAMStwofonts \else 
    \def\upi{\pi}
    \def\umu{\mu}
    \def\upartial{\partial}
  \fi
\fi

\title[]
  {A new direct method for measuring the Hubble constant from reverberating accretion
discs in active galaxies}
\author[S.J.~Collier et al.]
  {Stefan Collier$^1$, Keith Horne$^1$, Ignaz Wanders$^3$, and Bradley M. Peterson$^2$.
\\
$^1$University of St.~Andrews, School of Physics and Astronomy, North Haugh, St.~Andrews,
Fife, KY16 9SS. \\
$^2$Department of Astronomy, The Ohio State University, 174 West 18th Avenue, Columbus,
OH~43210. \\
$^3$Mercatorpad 4, Bus 401, 3000 Leuven, Belgium.}
\date{}
\pagerange{\pageref{firstpage}--\pageref{lastpage}}
\pubyear{1994}

\def\LaTeX{L\kern-.36em\raise.3ex\hbox{a}\kern-.15em
    T\kern-.1667em\lower.7ex\hbox{E}\kern-.125emX}

\begin{document}

\def\Msun{\ifmmode M_{\odot} \else $M_{\odot}$\fi}
\def\Hubble{\ifmmode {\rm km\,s}^{-1}\,{\rm Mpc}^{-1} \else km\,s$^{-1}$\,Mpc$^{-1}$\fi}
\def\arcsecpoint{$''\!.$}

\label{firstpage}

\maketitle

\begin{abstract}

We show how wavelength-dependent time delays between continuum flux
variations of AGN can be used to test the standard black hole-accretion
disc paradigm, by measuring the temperature structure $T(R)$ of the
gaseous material surrounding the purported black hole. Reprocessing of
high energy radiation in a steady-state blackbody accretion disc with $T
\propto R^{-3/4}$ incurs a wavelength-dependent light travel time delay
$\tau \propto \lambda ^{4/3}$. The International AGN Watch multiwavelength
monitoring campaign on NGC~7469 showed optical continuum variations
lagging behind those in the UV by about 1 day at 4800\AA\ and about 2 days
at 7500\AA. These UV/optical continuum lags imply a radial temperature
structure $T \propto R^{-3/4}$, consistent with the classical accretion
disc model, and hence strongly supports the existence of a disc in this
system. We assume that the observed time delays are indeed due to a
classical accretion-disc structure, and derive a redshift independent
luminosity distance to NGC~7469. The luminosity distance allows us to
estimate a Hubble constant of $H_{0} \left(\cos i / 0.7 \right)^{1/2} =
42\pm9\, \Hubble$. The interpretation of the observed time delays and
spectral energy distribution in the context of an accretion disc structure
requires further validation. At the same time, efforts to minimize the
systematic uncertainties in our method to derive a more accurate
measurement of $H_{0}$, e.g by obtaining an independent accurate
determination of the disc inclination $i$ or statistical average of a
moderate sample of active galaxies, are required. However, this remains a
promising new method of determining redshift-independent distances to
AGNs.

\end{abstract}


\section{Introduction}

The Hubble constant $H_{0}$ and deceleration parameter $q_{0}$ are
fundamental parameters in standard cosmology, measuring respectively the
rate at which the Universe is expanding and the rate at which that
expansion is impeded by the attractive force of gravity.  Moreover,
$H_{0}$ determines a size scale and age of the Universe, constrains the
baryonic density produced in the Big Bang, the amount of dark matter in
the Universe, and the epoch for galaxy and quasar formation in the early
Universe. A measurement of the deceleration parameter constrains the
geometry of the Universe. 

The value of the Hubble constant remains in
dispute after over half a century of intensive studies (Rowan-Robinson
1988, van den Bergh 1992, and de Vaucouleurs 1993). Broadly speaking there
are two distinct methods of calibrating distances to galaxies. The first
group of methods relies on accurate distances (parallaxes) to nearby
objects to calibrate a `distance ladder' extending to objects further
away. A recent successful example of this is the HST Key project (Freedman
et al.\, 1994 and Freedman et al 1997) that aims to measure the Hubble
constant with an accuracy of 10\% by using Cepheid variables as standard
candles to measure distances to the Virgo Cluster. Current estimates of
$H_{0}$ using this and similar methods (Tanvir et al.\, 1995)  are in the
range $\approx$ 60--90 \Hubble.  By accurately calibrating distances to
the Virgo cluster galaxies the `distance ladder' will be extended via
numerous secondary methods, including the D-$\sigma$ relation (Faber et
al.\, 1989), Tully-Fisher relationship (Tully \& Fisher 1977),
surface-brightness fluctuation method (Tonry \& Schneider 1988), and
supernova (Sandage et al.\, 1996 and Perlmutter et al.\, 1997) method. 

The
second group of methods does not require any calibration or progression
along a `distance ladder' but applies directly to the object concerned. 
Examples include the use of gravitational lens systems (Refsdael 1964 and
Kundic et al.\, 1997) and the Sunyaev-Zel'dovich effect ( Sunyaev \&
Zel'dovich 1980 and McHardy et al.\, 1990).  Current early estimates of
$H_{0}$ using these two methods are in the range $\approx$ 30--80 \Hubble. 
These direct methods, whilst generally more model dependent, give an
important check on `distance ladder' methods. 

AGNs are potentially
important cosmological probes because their high luminosity allows them to
be observed at large redshifts.  Correlations between the luminosity and
various emission line strengths and ratios (Baldwin 1977 and Kinney et
al.\, 1990) have been investigated for many years, but have not yet
allowed a consistently accurate inference of the distances of AGNs
independently of their redshifts.  A promising geometrical method that
uses proper motions and line-of-sight accelerations of water maser
emission in NGC~4258 yields an accurate distance of $6.4\pm0.9~{\rm Mpc}$
(Miyoshi et al. 1995).  In principle, a similar method may be employed to
infer distances to maser sources but not to much higher redshifts where
proper motions become too small. 

We propose a new method that utilizes the
relatively simple physics of light-travel time and blackbody radiation to
measure directly redshift-independent luminosity distances to AGNs, and
hence determine $H_{0}$. In \S{2} we discuss the theory of the method, and
show how wavelength-dependent time delays determine $T(R)$, and measure
$H_{0}$.  Therefore, wavelength-dependent time delays can be used to a)
provide `smoking gun' evidence of accretion disc structures in AGN and b) 
determine cosmological parameters, e.g $H_{0}$. The first application of
our method uses data from the International AGN Watch monitoring campaign
on NGC~7469, and is discussed in \S{3}. The systematic errors in our
method are presented in \S{4}, and \S{5} summarizes the main results of
the paper. 

\section{Theory}

A blackbody accretion disc illuminated by a central source has a radial
temperature profile $T(R)$ that is a non-linear combination of the surface
temperature due to viscous heat dissipation $T_{{\rm vis}}$ and that due
to irradiation $T_{{\rm irr}}$ of the disc, $T^{4} = T_{{\rm vis}}^{4} +
T_{{\rm irr}}^{4}$. Therefore $T(R)$ depends on both the geometry of the
accretion disc and the relative prominence of viscous heat dissipation and
irradiation effects.  For example, when $T(R)$ is determined by viscous
dissipation alone, a $T \propto (M\dot{M})^{1/4}\,R^{-3/4}$ structure
exists where $M$ is the mass of the black hole and $\dot{M}$ is the mass
accretion rate. An irradiating source, luminosity $L$, situated a height
$H_{{\rm x}}$ above the disc plane incurs a similar $T \propto (L H_{{\rm
x}})^{1/4}R^{-3/4}$ structure for $R \gg H_{{\rm x}}$, provided the disc
thickness $H \ll H_{{\rm x}}$.

The reprocessing hypothesis assumes that the UV/optical continuum
variations represent the response of gaseous material to variations in the
higher-energy continuum. The stringent upper limits, $< 0.3~{\rm day}$, on
time delays between the X-ray and UV variations in NGC~4151 (Edelson et
al.\, 1996) suggest that the variations in different wavebands must be
radiatively coupled (i.e., any possible time delays are due to
light-travel time effects), since, for example, viscous time scales are
much too long. Furthermore, the equivalent width of Fe K$\alpha$ at
6.4~KeV and the strength of Compton reflection observed at $> 10~{\rm
KeV}$ suggest that the majority of X-rays generated by an isotropic source
must be reprocessed by relatively cold ($< 10^{6}~{\rm K}$)  optically
thick gas, possibly that of an accretion disc (Pounds et al.\, 1990 and
George \& Fabian 1991). This has led to `ad hoc' models where the
higher-energy, e.g X-ray, source illuminates the accretion disc from
above.

We assume here that a similar mechanism must be operating, i.e. some
variable source of high-energy radiation in the vicinity of the disc axis
illuminates the disc and radiatively drives the UV/optical continuum
variations.  When high-energy radiation is emitted from the central
regions of the disc, a wave of heating propagates out at a speed $c$
arriving at radius $R$ after a mean time $\tau = R/c$. At this radius $R$
the temperature $T(R)$ rises slightly, thereby emitting more photons near
wavelength $\lambda = hc/kTX$ (where $X\approx$3--4 for blackbody
radiation). When we observe a time delay $\tau$ at wavelength $\lambda$
that in effect measures the radius $R=\tau c$ at which the disc has
temperature $T = hc / k\lambda X$. 

Assuming a temperature profile of the disc, $T=T_{0} (R/R_{0})^{-3/4}$,
the wavelength-dependent time delay is 
\begin{equation} 
\begin{array}{rl}
\tau (\lambda)= & 3.9~{\rm d} \left(\frac{T_{{0}}}{10^{4}~{\rm
K}}\right)^{4/3} \left(\frac{\lambda}{10^{4}~{\rm
\AA}}\right)^{4/3}\left(\frac{X}{4}\right)^{4/3}, 
\end{array}
\end{equation} 
where $T_{0}$ is the temperature of the disc at radius $R_{0}= 1$ light
day. Note that a disc with $T \propto R^{-3/4}$ predicts a $\tau \propto
\lambda ^{4/3}$ wavelength-dependent time delay, and an annulus of fixed R
in the disc responds with a range of time delays $\tau = (R/c)(1 \pm \sin
i)$, with $i$ the disc inclination. Hence, in principle, the width of the
time delay distribution at each wavelength determines $i$. The explicit
inclusion of $X$ in the above and following equations is for heuristic
purposes, since $X$ is not a free parameter or variable but is determined
by the blackbody model.

With $T(R)$ determined from the observed $\tau (\lambda)$, the 
predicted spectrum of the disc can be calculated straightforwardly by 
summing up the blackbody contributions from various disc 
annuli (Shakura and Sunyaev 1973).  The disc spectrum is given by
\begin{equation}
\begin{array}{rl}
f_{\nu} = & 11.2~{\rm Jy} \left(\frac{\tau}{{\rm days}}\right)^{2}
\left(\frac{D}{{\rm
Mpc}}\right)^{-2} \left(\frac{\lambda}{10^{4}~{\rm \AA}}\right)^{-3}
\left(\frac{X}{4}\right)^{-8/3} \cos i,
\end{array}
\end{equation}
where $D$ is the distance to the AGN and $i$ is the inclination of the
disc. Note that the classical thin disc spectrum, $f_{\nu} \propto
\lambda^{-1/3}$, is recovered since $\tau \propto \lambda ^{4/3}$,
hence $f_{\nu} \propto \tau ^{2} \lambda ^{-3} \propto \lambda 
^{-1/3}$. We note that $f_{\nu}$ is the distribution of flux with 
frequency $\nu$. 

The redshift-independent distance to the AGN is then derived to be
\begin{equation}
\begin{array}{rl}
D = & 3.3~{\rm Mpc} \left(\frac{\tau}{{\rm days}}\right)
\left(\frac{\lambda}{10^{4}~{\rm \AA}}\right)^{-3/2} \left(\frac{f_{\nu}/
\cos i}{{\rm Jy}}\right)^{-1/2} \left(\frac{X}{4}\right)^{-4/3}. 
\end{array}
\end{equation}
By inserting observed values of $\tau (\lambda)$ and $f_{\nu}$ into the
above equation, we determine a redshift-independent distance to the
object. Hubble's constant, $H_{0}=cz/D$, is then,
\begin{equation}
\begin{array}{rl}
H_{0} = & 89.6~\frac{{\rm km}~{\rm s}^{-1}}{{\rm Mpc}} 
        \left( \frac{ \lambda }{ 10^{4}\AA } \right)^{3/2 }
        \left( \frac{ z }{ 0.001 } \right)
        \left( \frac{ \tau }{ {\rm day} } \right)^{-1}
        \left( \frac{ f_{\nu} / \cos i }{ {\rm Jy} } \right)^{1/2 }
\\  &  \left( \frac{X}{4} \right)^{4/3},
\end{array}
\end{equation}
where $z << 1$ is the redshift of the AGN.

\section{Application of Method to NGC~7469 Monitoring Data}

To apply our new method we use data from a seven-week International AGN
Watch (Alloin et al.\, 1994)  multiwavelength monitoring campaign on
NGC~7469, $z=0.0164$, which showed optical continuum variations lagging
behind those in the UV by about 1 day at 4800\AA\ and about 2 days at
7500\AA\ (Wanders et al.\, 1997 and Collier et al.\, 1998). These
UV/optical continuum lags have been shown to be statistically significant
at no less than 97\% confidence (Peterson et al.\, 1998). The measured
time delay between the flux variations at wavelength $\lambda$ and those
at wavelength 1315\AA\ is shown in the top panel of Figure 1.\ The
observed time delays are encompassed by the $\tau \propto \lambda ^{4/3}$
predictions, shown as dashed and dotted lines, for irradiated blackbody
accretion disc models with $T = T_{0} (R/R_{0})^{-3/4}$ structure, with
$T_{0}^{{\rm min}} \approx 6500~{\rm K}$ and $T_{0}^{{\rm max}} \approx
7700~{\rm K}$ for faint and bright states respectively. The observed
delays increase above the overall trend at wavelengths near emission
lines.  Here there is a mix of continuum and line flux, the lines
responding with larger delays than the continuum light. The horizontal
axes in Fig.\ 1 give for each wavelength $\lambda$ the corresponding
temperature $T$ in the disc.  The temperature is calculated assuming
$T=hc/k\lambda X$, with $X=3.89$ being appropriate for blackbody discs
with $T \propto R^{-3/4}$ as indicated by numerical simulations. 
\begin{figure*}
\centerline{\epsfig{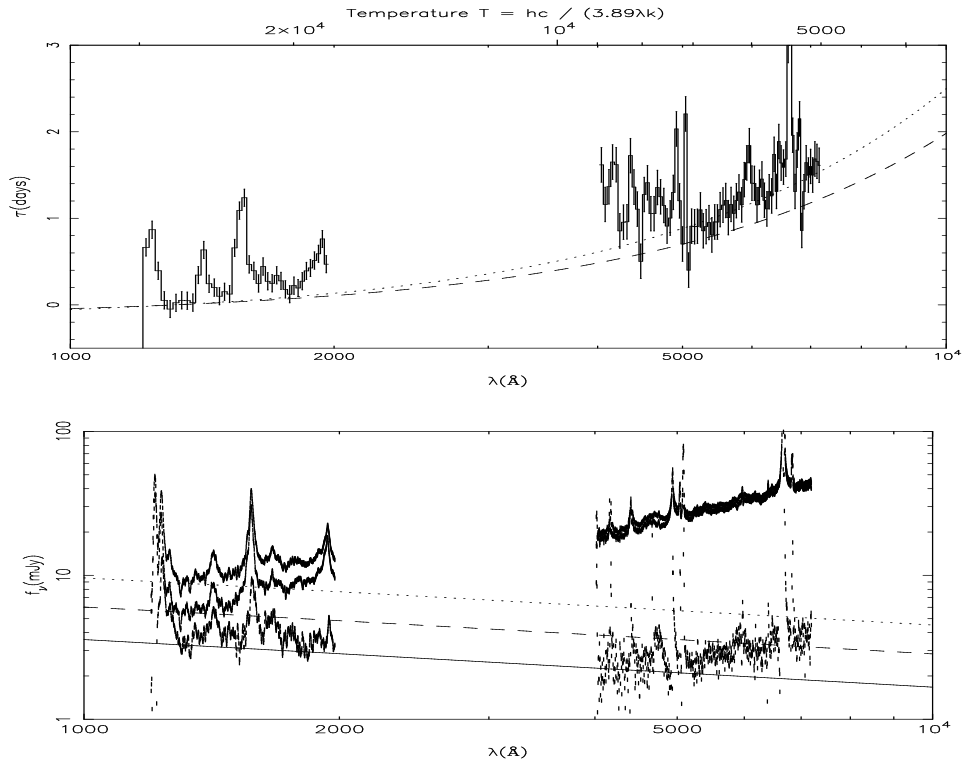}}
\vspace{10pt}
\caption{Top Panel: The predicted time delays for the irradiated accretion
disc model are compared with the observed time delays, measured relative
to 1315\AA. The time delay increases above the overall trend near the
emission lines because here there is a mix of continuum and line flux. The
lines respond with longer time delays and this results in a larger net
delay. Bottom panel: Model spectra for an irradiated accretion disc
(assuming $H_{0} = 42\,\Hubble$ and $i=45^{\circ}$) are compared with
observed spectra of NGC~7469 from the 1996 AGN Watch multiwavelength
monitoring campaign. See text for full details.}
\label{}
\end{figure*} 

The
bottom panel of Fig.\ 1 compares the de-reddened UV/optical maximum,
minimum and difference spectra of the NGC~7469 monitoring campaign with
predicted spectra for $H_{0}=42~\Hubble$ ($D \approx 117~{\rm Mpc}$) based
on the irradiated accretion disc model inclined at $i=45^{\circ}$ to the
observer's line-of-sight. The dotted and dashed lines represent model
spectra for bright and faint states of the irradiated accretion disc, and
define maximum and minimum temperatures, $T_{0}^{{\rm max}}\approx
7700~{\rm K}$ and $T_{0}^{{\rm min}} \approx 6500~{\rm K}$, at radius
$R_{0}=1$ light day respectively.  The brightest and faintest spectra seen
during the NGC~7469 campaign are both much redder than the predicted disc
spectra. This is caused by contamination of the observed spectra by a red
starlight component from the host galaxy.  We therefore consider these
spectra to be upper limits to the spectrum of the active nucleus. The
solid line represents the predicted difference spectrum between the bright
and faint states of the disc.  This agrees approximately with the
difference spectrum between the brightest and faintest spectra recorded in
the NGC~7469 AGN Watch campaign. The difference spectrum, which cancels
any starlight contamination, gives a lower limit to the nuclear spectrum. 

The results shown in Fig.\,1 demonstrate that the observed variability in
the continuum spectrum of NGC~7469 is in approximate agreement with a
blackbody disc, $T \propto R^{-0.75}$, and $H_{0}\left(\cos i
\right)^{1/2} = 35\pm6\, \Hubble$. The errorbars reflect statistical
uncertainties in the measured fluxes $f_{\nu}$, time delays $\tau$ and
redshift $z$, e.g.\, equation $(4)$. The point-to-point scatter in the
continuum time delays allows us to assign a 10\% uncertainty to the time
delay measurements. The lower limit to the flux of the nucleus, is $\Delta
f_{\nu} = 3.5\pm1.0~{\rm mJy}$ at $\lambda = 1315$\AA. The uncertainty in
the redshift of NGC~7469 ($z=0.0164$), induced by assuming $\pm 300~{\rm
km}~{\rm s}^{-1}$ peculiar velocities (Lynden-Bell et al.\, 1988), is 6\%. 
Our 10\% uncertainty in $\tau$, 29\% and 6\% uncertainties in $\Delta
f_{\nu}$ and $z$ respectively, result in a 18\% uncertainty in
$H_{0}\left(\cos i \right)^{1/2}$.

\section{Systematic Errors}

Our estimate of $H_{0}$ is subject to several systematic errors. The AGN
spectrum is diminished and reddened by intervening dust.  Reddening
estimates derived from pointed 21cm observations give $E(B-V) \approx
0.074$--$0.096$ (Elvis et al.\, 1989 and Lockman \& Savage 1995). Other
estimates based on using the 2200\AA\ dust absorption feature give $E(B-V) 
\approx 0.14$ (Westin 1985). We have corrected our spectra using
$E(B-V)=0.14$. The host-galaxy contamination can be estimated from
off-nuclear observations of the host galaxy, e.g using {\it HST} or ground
based adaptive optics. In the difference spectra the host-galaxy
contamination is negligible. The red slope of the mean spectrum is due to
contamination by starlight from the host galaxy, which contributes at
least $\approx$ 40\% at 5400\AA\ in a circular $10$ arcsecond radius
aperture (Malkan \& Filippenko 1983). Welsh et al.\, 1998 present
contemporaneous HST observations of NGC~7469 and estimate a percentage
host
galaxy contamination at 7400\AA\ of $\approx$ 80\% in a 10$''$ $\times$ 
16\arcsecpoint8 aperture.

A systematic error
arises from uncertainty in $\left(\cos i \right)^{1/2}$. However, the
inclination uncertainty is not a major obstacle. According to unified
schemes (Antonucci 1993 and Hes et al.\, 1993), Seyfert 1 galaxies, in
which we see the broad emission line region (BLR), have $i < 60^{\circ}$,
while Seyfert 2 galaxies, in which the BLR is obscured by a dusty torus,
have $i > 60^{\circ}$. For $ i < 60^{\circ}$, $\left(\cos i \right)^{1/2}
> 0.7$. By adopting $i=45^{\circ}$, $\left(\cos i \right)^{1/2}=0.84$, we
commit a maximum error of $\pm\, 17\%$, and an RMS error of $\pm\, 11 \%$.
Averaging over 10 objects could reduce this by a factor $ \sqrt {10}$ to
$\pm\, 3.5\%$.  With $\left(\cos i \right)^{1/2}=0.84\pm0.1$ we find
$H_{0}=42\pm9\, \Hubble$. We may be able to reduce the uncertainty in
$\left(\cos i \right)^{1/2}$ if we can measure $i$ independently rather
than averaging over the full range of $i$.  Our echo mapping method in
principle allows us to derive the inclination $i$, because the width of
the time delay distribution at each wavelength is a function of $i$. This
method may be applied, in future, to NGC~7469 and other Seyfert 1
galaxies. However, it is likely to be a difficult task. Fits to the
profile of the X-ray Fe\,K$\alpha$ line in MCG-6-30-15 (Tanaka et al.\,
1995) yielded a disc inclination $i$ of $30\pm3^{\circ}$. Similar
observations of an X-ray Fe\,K$\alpha$ line in NGC~7469 could therefore
measure $i$ with about 10\% accuracy. It may also be possible to determine
$i$ from polarization measurements. Therefore there are good prospects for
measuring inclinations of individual AGNs. 

The blackbody model is a source
of systematic uncertainty.  Our value $X=3.89$ relies on the assumption
that the changes in the UV/optical continuum can be modelled as
irradiation of a blackbody disc. This is justified by the approximate
agreement of the predicted $f_{\nu} \propto \lambda^{-1/3}$ spectrum with
the observed difference spectrum. This can be further investigated by
considering models of the vertical structure and the emitted spectra of
irradiated accretion disc atmospheres (Sincell \& Krolik 1997). We expect
that to first order our method of measuring distances should be
insensitive to limb darkening (Hubeny et al.\, 1997).  First, the
irradiation of the disc will flatten the temperature versus optical depth
relationship in the atmosphere and conspire to reduce the limb darkening
effect. Second, the lower temperatures observed at high inclinations will
change the apparent $T(R)$ profile, but the same blackbody relationship
will describe the surface brightness distribution. Hence, while our $T(R)$
profile is sensitive to limb darkening the inferred surface brightness and
subsequent distance is not. 

Another source of systematic uncertainty is
the source geometry and nature of the continuum variations. The
reprocessing geometry we have considered might be completely ruled out by
the recent X-ray observations of NGC~7469 (Nandra et al.\, 1998). The 2-10
KeV X-ray variations are poorly correlated or uncorrelated with the
UV/optical variations described here. Either (a) the accretion disc sees
different X-ray variations than we do, (b) the UV/optical variations are
driven by another unobserved part of the spectrum, e.g. the extreme UV, or
(c) the model geometry is completely wrong. However, observations of time
delays between different wavebands and knowledge of the spectral energy
distribution of sources will allow us to determine the importance or
irrelevance of these various systematic effects.

\section{Summary}

Our Hubble constant estimate of $H_{0}=42\pm9\, \Hubble$ is not consistent
with the independent `distance ladder' estimates of $H_{0}=80\pm17\,
\Hubble$ by Freedman et al.\, 1994, and $H_{0}=69\pm8\, \Hubble$ by Tanvir
et al.\, 1995.  However, the 22\% accuracy of our Hubble constant estimate
compares favourably with the 21\% and 12\% uncertainties of the Hubble
constant estimates reported by Freedman and Tanvir respectively. For
comparison, the direct method utilizing gravitational lenses give, for
example, Hubble constant estimates of $H_{0}=42\pm6\, \Hubble$ (Schecter
et al.\, 1997), $H_{0}=53^{+10}_{-7}\, \Hubble$ (Courbin et al.\, 1997),
and $H_{0}=64\pm13\, \Hubble$ (Kundic et al.\, 1997). The
Sunyaev-Zel'dovich method gives $H_{0}=47^{+23}_{-15}\, \Hubble$ (Hughes
and Birkinshaw 1998 and references therein).  Any apparent discrepancy in
the Hubble constant estimates from `distance ladder' and direct methods
will need investigation, which in turn requires statistically significant
samples of the methods to be compared. 

Our result based on observations not specifically designed to measure
$H_{0}$ can be improved upon. A continuous 2-3 month multiwavelength,
multi-telescope monitoring campaign on a sample of AGNs will make this
method a serious competitor with the established methods of measuring
$H_{0}$. Principally, the time delay measurements can be constrained to
better than 5\%. There are good prospects for measuring inclinations of
individual AGNs with high accuracy.  The starlight contamination can be
estimated and corrected for as already described, allowing an accurate
determination of the nuclear spectrum to better than 10\%.  Finally, a
statistical average of individual $H_{0}$ measurements will reduce our
final uncertainty, and we expect a future optimally designed experiment to
determine the Hubble constant with $\le 10\%$ accuracy.

To conclude, we identify the {\it variable} component of the de-reddened
UV and optical continuum fluxes as a lower limit to the nuclear spectrum. 
For NGC~7469 we note that this agrees approximately with the $f_{\nu}
\propto \lambda^{-1/3}$ spectrum predicted for a blackbody accretion disc
with a $T \propto R^{-3/4}$ structure.  At the same time the
wavelength-dependence of the observed time delays, $\tau \propto \lambda
^{4/3}$, is consistent with $T \propto R^{-3/4}$.  The concurrence of
these two independent lines of evidence strongly supports the notion of a
standard blackbody accretion disc in NGC~7469, and strengthens the
evidence (Shields 1978, Malkan et al.\, 1982, and Tanaka et al.\, 1995)
for accretion discs in AGN. Using the {\it variable} component of the
continuum fluxes we find $H_{0} \left(\cos i / 0.7 \right)^{1/2} =
42\pm9\,\Hubble$. The interpretation of the observed time delays and
spectral energy distribution in the context of an accretion disc structure
requires further validation.  However, analysis of the observed variable
spectrum and wavelength-dependent time delays along the lines outlined
above yields redshift-independent luminosity distances to AGNs. This opens
up a new route to $H_0$ and by extension to fainter objects at $z \sim 1$,
$q_0$.

\subsection*{Acknowledgements}

SC acknowledges support from a PPARC graduate studentship to the School of
Physics \& Astronomy at St.~Andrews. BMP acknowledges support from NASA
and NSF.

\end{document}